\documentclass[aps,prl,reprint,showpacs,superscriptaddress,nobibnotes,nofootinbib]{revtex4-1}
\usepackage{graphicx}
\usepackage{amssymb} 
\usepackage{amsmath}
\begin{document}

\title{New limit on Lorentz and \textit{CPT} violating neutron spin interactions using a free precession $^{3}$He-$^{129}$Xe co-magnetometer}
\author{F. Allmendinger}
\email{Corresponding author: allmendinger@physi.uni-heidelberg.de}
\affiliation{Physikalisches Institut, Ruprecht-Karls-Universit\"{a}t, 69120 
Heidelberg, Germany}

\author{W. Heil}
\affiliation{Institut f\"{u}r Physik, Johannes Gutenberg-Universit\"{a}t, 
55099 Mainz, Germany}

\author{S. Karpuk}
\affiliation{Institut f\"{u}r Physik, Johannes Gutenberg-Universit\"{a}t, 
55099 Mainz, Germany}

\author{W. Kilian} 
\affiliation{Physikalisch-Technische Bundesanstalt Berlin, 10587 Berlin, Germany}

\author{A. Scharth}
\affiliation{Institut f\"{u}r Physik, Johannes Gutenberg-Universit\"{a}t, 
55099 Mainz, Germany}

\author{U. Schmidt}
\affiliation{Physikalisches Institut, Ruprecht-Karls-Universit\"{a}t, 69120 
Heidelberg, Germany}

\author{A. Schnabel} 
\affiliation{Physikalisch-Technische Bundesanstalt Berlin, 10587 Berlin, Germany}

\author{Yu. Sobolev} 
\affiliation{Institut f\"{u}r Physik, Johannes Gutenberg-Universit\"{a}t, 55099 Mainz, Germany}

\author{K. Tullney}
\affiliation{Institut f\"{u}r Physik, Johannes Gutenberg-Universit\"{a}t, 
55099 Mainz, Germany}

\date{\today}
\begin{abstract}
We report on the search for a \textit{CPT} and Lorentz invariance violating coupling of the $^3$He and $^{129}$Xe nuclear spins (each largely determined by a valence neutron) to background tensor fields which permeate the universe. Our experimental approach is to measure the
free precession of nuclear spin polarized $^3$He and $^{129}$Xe atoms in a homogeneous magnetic guiding field of about 400 nT using LT$_C$ SQUIDs as low-noise magnetic flux detectors. As the laboratory reference frame rotates with respect to distant stars, we look for a sidereal modulation of the Larmor frequencies of the co-located spin samples. As a result we obtain an upper limit on the equatorial component of the background field interacting with the spin of the bound neutron $\tilde{b}^{\text{n}}_{\bot} < 6.7 \cdot 10^{-34}$ GeV (68\%\,\,C.L.). Our result improves our previous limit (data measured in 2009) by a factor of 30 and the world's best limit by a factor of 5.
\end{abstract}


\maketitle

\begin{figure}
\includegraphics{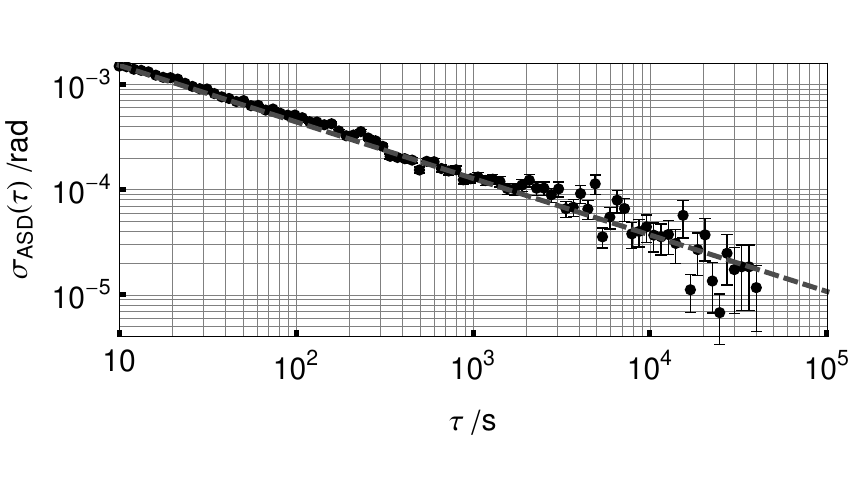}
\caption{Allan Standard Deviations (ASD) of the residual phase noise of a single run ($j=6$). The total observation time was $T=90000$ s. With increasing integration times $\tau$ the uncertainty in phase decreases as $\propto \tau^{-\frac{1}{2}}$ indicating the presence of white phase noise.}
\label{fig:ASD}
\end{figure}

A great number of laboratory experiments have been designed to detect diminutive violations of \textit{CPT} and Lorentz invariance. Unlike Michelson-Morley type experiments \cite{Michelson, Eisele, Herrmann}, Hughes-Drever experiments \cite{Hughes,Drever} test the isotropy of the interactions of matter itself. Searches for an anomalous spin coupling to a relic background field which permeates the universe have been performed with electron and nuclear spins with increasing sensitivity \cite{Prestage,Brown,Lamoreaux,Lamoreaux2,Chupp,Berglund,Bear,Bear2,Phillips,Cane,Heckel,Altarev,Peck}. The theoretical framework presented by A. Kosteleck\'y and colleagues parametrizes the general treatment of \textit{CPT} and Lorentz invariance violating (LV) effects in a Standard Model Extension (SME) \cite{Colladay,Kostelecky,Kostelecky2}. The SME was conceived to facilitate experimental investigations of Lorentz and \textit{CPT} symmetry, given the theoretical motivation for violation of these symmetries. Although Lorentz-breaking interactions are motivated by models such as string theory \cite{Kostelecky2,Ellis}, loop quantum gravity \cite{Sudarsky,Nicolai,Gambini,Crichigno}, etc., the low-energy effective action appearing in the SME is independent of the underlying theory. The SME contains a number of possible terms that couple to the spins of fundamental Standard Model particles like the electron, or composite particles like the proton and (bound) neutron. These terms are small due to Planck-scale suppression ($M_p$), and in principle are measurable in experiments by detecting tiny energy shifts of order $\Delta E^{(n)}\sim (\frac{m_w}{M_p})^n\cdot m_w$, where the low energy scale is set by the mass $m_w$ of the particle. Since $n=1$ is largely ruled out by present experiments \cite{KosteleckyTables}, tuning the measurement sensitivity to second order effects ($n=2$) in Planck scale suppression is the current challenge\footnote{For the neutron ($m_n =939$ MeV) this is $\Delta E^{(2)} \approx 10^{-38}$ GeV.}. To determine the leading-order effects of a LV potential $V$, it suffices to use a non-relativistic description for the particles involved given by
\begin{equation} \label{equ:SME1} 
V=-\tilde{b}_{J}^{w} \cdot \sigma _{J}^{w} \,\,\,\,\,\,\,\, (\text{with  } J~=~X,Y,Z~;~~~ w~=~e, p, n)~,
\end{equation} 
which can be interpreted as a coupling of the electron, proton or neutron spin $\sigma_{J}^{w}$ to a hypothetical background field $\tilde{b}_{J}^{w} $.
The most sensitive tests so far were performed on the bound neutron using a $^{3}$He-$^{129}$Xe Zeeman maser \cite{Bear,Bear2}, a $^{3}$He-$^{129}$Xe co-magnetometer \cite{Gemmel2} based on free spin precession, and a K-$^{3}$He co-magnetometer \cite{Brown}. The latter one so far gave the highest energy resolution of any spin anisotropy experiment.\\
The experiment described here is a continuation of our year 2009 measurements \cite{Gemmel2} with some essential improvements (see below) and is based on the detection of freely precessing nuclear spins of  polarized $^3$He and $^{129}$Xe gas with SQUIDs as low-noise magnetic flux detectors. Like in \cite{Bear,Bear2}, we search for sidereal variations of the frequency of co-located spin species while the Earth and hence the laboratory reference frame rotates with respect to distant stars. The $^3$He-$^{129}$Xe co-magnetometer has been described in detail in \cite{Gemmel, Gemmel2, Heil,Tullney}. Briefly, we used a LT$_C$ DC-SQUID magnetometer system inside the strongly magnetically shielded room BMSR-2 at PTB Berlin \cite{Bork} (latitude $\Theta=52.52^\circ$). A homogeneous magnetic guiding field $B_0$ of about 400 nT was provided by two square coil pairs which were arranged perpendicular to each other in order to manipulate the sample spins, e.g., $\pi$/2 spin-flip by non-adiabatic switching. 
The $^3$He and $^{129}$Xe nuclear spins were polarized outside the shielding by means of optical pumping. Low-relaxation spherical glass vessels ($R=5$ cm) were filled with the polarized $^3$He and $^{129}$Xe gases and placed below the Dewar housing the SQUID sensors, which detect the sinusoidal change in magnetic flux due to the spin precession of the gas atoms. In order to obtain a high common mode rejection ratio, gradiometric sensor arrangements (four in total) were used. Typically, the optimum conditions in terms of long transverse relaxation times ($T_2^*$) and high signal-to-noise ratio (SNR) were met at a gas mixture with pressures of ($^{3}$He, Xe (91\% $^{129}$Xe), N$_2$)=(3, 5, 25) mbar. Nitrogen was added as buffer gas to suppress spin-rotation coupling in bound Xe-Xe van der Waals molecules \cite{Chann,Anger}. In total 7 measurement runs  ($j$ = 1, \dots ,7) with free spin precession were performed, each lasting about one day. For the first three runs resp. for the last four ones, the magnetic guiding field pointed at an angle of $\rho_1=208^{\circ}$ resp. $\rho_2=73^{\circ}$ to the north-south direction. 
The recorded signal is a superposition of the $^{3}$He and $^{129}$Xe precession signals at Larmor frequencies $\omega_{\text{He}} =\gamma_{\text{He}} \cdot B_{0} \approx 2\pi \cdot 13.0$\:Hz and $\omega_{\text{Xe}} =\gamma_{\text{Xe}} \cdot B_{0} \approx 2\pi \cdot 4.7$\:Hz.\footnote{$\frac{\gamma_{He}}{\gamma_{Xe}}\approx 2.75408159(20)$ \cite{Codata,Pfeffer}.}
Analogue to other experiments with high precision in frequency or phase determination \cite{Hoyle2} phases of sub-data sets were analyzed: The data of each run was divided into sequential time intervals of $\tau$~=~3.2\:s. The number $N_{j}$ of obtained sub-data sets laid between 20000 and 28000 corresponding to observation times $T_{j}$ of coherent spin precession in the range of 18\:h to 25\:h. For each sub-data set ($i$) a $\chi^{2}$ minimization was performed, using the fit function 
\begin{align} \label{GrindEQ__4_} 
\nonumber A^{i} (t) = &A_{\text{He}}^{i} \cdot \sin \left(\omega _{\text{He}}^{i} t\right)+
B_{\text{He}}^{i} \cdot \cos \left(\omega _{\text{He}}^{i} t\right)+ \\
\nonumber & A_{\text{Xe}}^{i} \cdot \sin \left(\omega _{\text{Xe}}^{i} t\right)+
 B_{\text{Xe}}^{i} \cdot \cos \left(\omega _{\text{Xe}}^{i} t\right)+ \\
&  c_{0}^{i} +c_{\text{1}}^{i} \cdot t
\end{align} 
\noindent with a total of 8 fit parameters. Within the relatively short time intervals, the term $c_{0}^{i} + c_{\text{1}}^{i} \cdot t$ represents the adequate parameterization of the SQUID gradiometer offset showing a small linear drift due to the elevated 1/$f$ noise at low frequencies ($<$ 1\:Hz) \cite{Gemmel}. On the other hand, the chosen time intervals are long enough to have a sufficient number of data points (800) for the $\chi^{2}$ minimization. The sum of sine and cosine terms are chosen to have linear fitting parameters (except $\omega _{\text{He}}$ and $\omega _{\text{Xe}}$) with orthogonal functions. The reduced $\chi^{2}$ ($\chi^{2}$/d.o.f.) of most sub-data sets is close to $1$ which is consistent with the assigned uncertainty to each data point of $\pm$\:16\:fT (1$\sigma$), typically. 
The initial signal amplitudes were up to 10\:pT (Xe) and 20\:pT (He).
For each sub-data set we finally obtain numbers for the respective fit parameters $A^i_{He}, B^i_{He},A^i_{Xe}, B^i_{Xe}, \omega_{\text{He}}^{i}, \omega_{\text{Xe}}^{i}$ including their uncertainties. The sub-data set phases are then calculated using
\begin{equation} \label{equ:arctan} 
\varphi^{i} =\arctan \, \, (B^{i} /A^{i})~.
\end{equation}
The accumulated phases for helium and xenon at the times $t=i\cdot\tau$ are then determined by adding appropriate multiples of $2\pi$. After one day of measurement the accumulated phase for $^3$He is $\Phi_{He}\approx7\cdot10^6$ rad, e.g.. In the next step the weighted phase difference $\Delta \Phi \left(t=i\tau\right)$ is calculated with
\begin{equation} \label{equ:wphasediff} 
\Delta \Phi =\Phi _{He} -\frac{\gamma _{He} }{\gamma _{Xe} } \cdot \Phi _{Xe}.         
\end{equation} 
By that measure the Zeeman term is eliminated and thus any dependence on fluctuations and drifts of the magnetic guiding field, i.e. $\Delta\Phi=const.$ .
However, non-magnetic spin interactions, like the one of Eq. (\ref{equ:SME1}), do not drop out and we expect a sidereal modulation of the weighted phase difference, i.e., $\Delta\Phi\propto \sin(\Omega_{s}\cdot t + \varphi)$,\footnote{$\Omega_{s}=2\pi/(23^h:56^{min}:4.091^s)$  is the angular frequency of the sidereal day.} that allows to distinguish between a normal magnetic field and an anomalous field coupling. 
On a closer inspection, the effect of Earth's rotation (i.e. the rotation of the SQUID detectors with respect to the precessing spins) is not compensated by co-magnetometry as well as frequency shifts due to the Ramsey-Bloch-Siegert (RBS) shift \cite{Bloch, Ramsey}. The latter one gives the shift in Larmor frequency $\omega_L$ due to a rotating field with amplitude $B_1$ and frequency $\omega_D$, which, related to our case, is generated by the precessing magnetization of the polarized gas:
\begin{equation}
\delta\omega_{\text{RBS}}(t)=\pm\left(\sqrt{\Delta\omega^2+\gamma^2B_1^2(t)}-\Delta\omega\right)
\end{equation}
with $\Delta\omega=|\omega_L-\omega_D|$. The plus sign applies to $\frac{\omega_D}{\omega_L}<1$, the minus sign to $\frac{\omega_D}{\omega_L}>1$, respectively.
Two effects contribute to the RBS shift and have to be taken into account, i.e., cross-talk (ct) and self-shift (ss). i) Concerning cross-talk, i.e., the coupling of the $^{3}$He and $^{129}$Xe magnetic moments among each other, $\Delta\omega\gg \gamma B_1$ holds. For a precessing spherical spin sample with magnetization $\vec{M}(t)$ we have $|\vec{B_1}(t)|=\frac{2\mu_0}{3}|\vec{M}(t)|=|\vec{B_1}(0)|\cdot\exp(-t/T_2^*)$ for the magnetic field produced inside the sample cell \cite{Jackson} and expressions for the weighted phase difference (Eq. (\ref{equ:wphasediff})) can be calculated by integrating over time:
\begin{eqnarray}
\nonumber \Delta\Phi_{\text{RBS}}^{\text{ct}}(t)&=&F_{He}\cdot e^{-\frac{2\cdot t}{T_{2, Xe}^*}}-F_{Xe}\cdot e^{-\frac{2\cdot t}{T_{2, He}^*}}\\
\nonumber &=&\frac{-\gamma_{He}^2B^2_{1,Xe}(0)\cdot T_{2, Xe}^*}{4\Delta\omega}\cdot e^{-\frac{2\cdot t}{T_{2, Xe}^*}}\\
&&+\frac{\gamma_{He}\gamma_{Xe}B^2_{1,He}(0)\cdot T_{2, He}^*}{4\Delta\omega}\cdot e^{-\frac{2\cdot t}{T_{2, He}^*}}
\end{eqnarray}
The magnetic field inside the sample cell $B_1(t)$ can be determined by analyzing the signal of the SQUID magnetometers which directly measure the magnetic dipole field outside the spherical sample cell at their respective positions. The uncertainty in the determination of $B_1(0)$ is in the order of $5\%$, resulting in uncertainties of $\Delta\Phi_{\text{RBS}}^{\text{ct}}$ of about $10\%$. Taking into account the uncertainties on $F_{\text{He}}^{(j)}$ and $F_{\text{Xe}}^{(j)}$, the final fitting procedure was not a basic $\chi^2$-minimization, but a maximization of the likelihood $L$ including the Gaussian probability distributions for $F_{\text{He}}^{(j)}$ and $F_{\text{Xe}}^{(j)}$ \cite{Kay}. In contrast, the characteristic time constants $T_{2}^*$ of the exponential decay of the precession signals as well as the Larmor frequencies could be determined with high precision, so that they enter as fixed values into the fitting procedure.
ii) For the self-shift, i.e., coupling of the precessing magnetic moments of the same spin species, $\Delta\omega\ll\gamma B_1$ holds  with $\delta\omega_{\text{RBS}}(t)\propto\gamma B_1(t)\propto \exp{(-\frac{t}{T_2^*})}$ and for the corresponding expression of the weighted phase difference we get 
\begin{equation}
\Delta\Phi_{\text{RBS}}^{\text{ss}}(t)=E_{He}\cdot e^{-\frac{t}{T_{2, He}^*}}-E_{Xe}\cdot e^{-\frac{t}{T_{2, Xe}^*}}
\end{equation}
$B_1(t)$, in turn, depends on the source strength and thus must show the time dependence of the signal amplitude ($\propto\exp(-t/T_2^*)$). However, the proportionality factor strongly depends on the field gradients across the sample cell, the resulting diffusion coefficients for $^{3}$He and $^{129}$Xe in the gas mixture, and the geometry of the sample cell \cite{Gemmel}. During the duration of a single run ($j$), these parameters are sufficiently constant, so that only the time dependence of the signal amplitude enters. Since we have no precise enough model to calculate the amplitudes of $\Delta\Phi_{\text{RBS}}^{\text{ss}}$, $E_{He}$  and $E_{Xe}$ must be kept as free fit parameters. 
If there is no sidereal variation of the $^{3}$He and $^{129}$Xe frequencies induced by LV couplings, then the time dependence of the weighted phase difference can be described best by the fit model
\begin{align} \label{eqn:fitmodel1} 
\Delta\Phi_{\text{c}}(t)&=&\left\{\begin{array}{l} {\Delta\Phi _{\text{d}}^{(1)}(t)\, \, \, \, \mbox{for}\, \, \, t_{0,1} \le t\le (t_{0,1} +N_{1} \cdot \tau)} \\ 
{\,\,\,\,\,\,\,\,\,\,\,\,\,\,\,\,\,\,\,\,\,\,\,\,\vdots }\\
{\Delta\Phi _{\text{d}}^{(7)}(t)\, \, \, \, \mbox{for}\, \, \, t_{0,7} \le t\le (t_{0,7} +N_{7} \cdot \tau)}\\
 \end{array}\right.
\end{align}
with
\begin{eqnarray} \label{eqn:fitmodel2} 
\Delta\Phi_{\text{d}}^{(j)}\left(t\right)&=&\Phi_{0}^{(j)}+\Delta\omega_{\text{lin}}^{(j)}\cdot\left(\,t-t_{0,j}\right) \nonumber \\
&&+E_{\text{He}}^{(j)}\cdot e^{\frac{-\left(\,t-t_{0,j}\right)}{T_{2,\text{He}}^{*\, (j)}}}-E_{\text{Xe}}^{(j)}\cdot e^{\frac{-\left(\,t-t_{0,j}\right)}{T_{2,\text{Xe}}^{*\, (j)}}} \nonumber\\
&&-F_{\text{Xe}}^{(j)}\cdot e^{\frac{-2\left(\,t-t_{0,j}\right)}{T_{2,\text{He}}^{*\, (j)}}}+F_{\text{He}}^{(j)}\cdot e^{\frac{-2\left(\,t-t_{0,j}\right)}{T_{2,\text{Xe}}^{*\, (j)}}}\nonumber\\
\end{eqnarray}
\begin{figure}
\begin{flushright}
\includegraphics{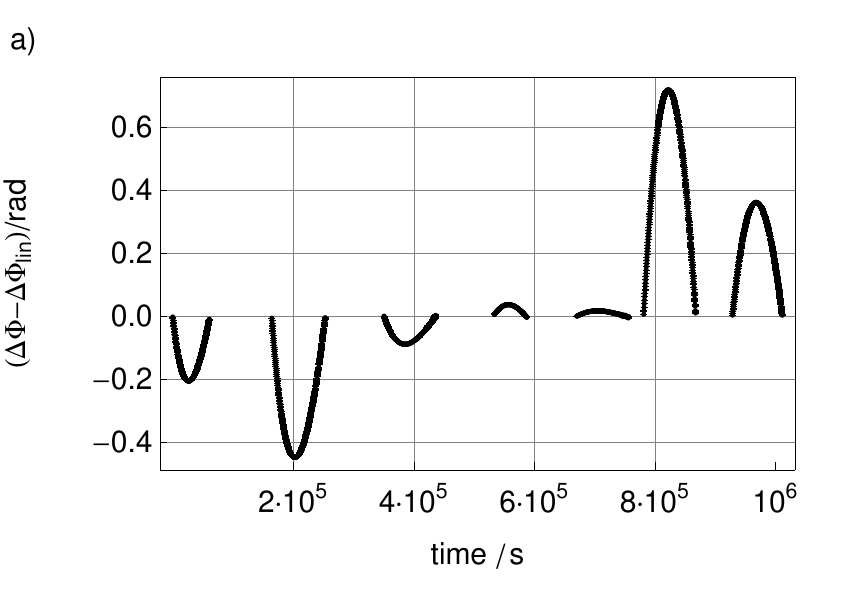}
\includegraphics{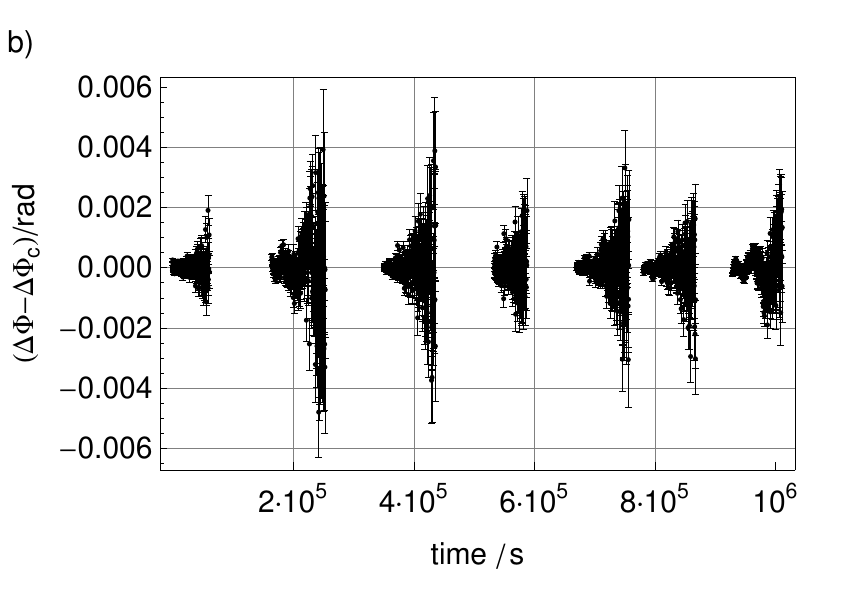}
\end{flushright}
\caption{(a) Weighted phase difference $\Delta\Phi$ (data-bin: 320 s) after subtraction of the linear terms $\Delta\Phi_{\text{lin}}$ in Eq. (\ref{eqn:fitmodel2}). The remaining parabolic shaped structure is the contribution of the RBS-shift (in particular the self-shift). (b)  Phase residuals after subtraction of the entire fit-model $\Delta\Phi_c$ according to Eqs. (\ref{eqn:fitmodel1}) and (\ref{eqn:fitmodel2}). The time evolution of the phase noise is caused by the exponential decay of the signal amplitudes. Note that the phase noise is much less than the symbol size in (a).
}
\label{fig:residuals}
\end{figure}
$t_{0,j}$ is the starting time of the corresponding run ($j$) (with $t$=0 at 15:35 UT on March 7th, 2012) \cite{siderealphase}. As for the RBS amplitudes, it is generally found that $|E_{He(Xe)}^{(j)}|\gg |F_{He(Xe)}^{(j)}|$.\footnote{In \cite{Gemmel2} the contribution of the cross-talk could be neglected due to the lower SNR ($\sim B_1$).} Fitting the data for the weighted phase difference to Eq. (\ref{eqn:fitmodel2}) and subtracting the fit function, results in the phase residuals as shown in Fig.~\ref{fig:residuals}b. Due to the exponential decrease of the  $^{3}$He and $^{129}$Xe signal amplitudes (in particular the xenon amplitude) the residual phase noise rises in time ($\sigma_{res}\propto \exp(t/T_{2, \text{Xe}}^{*})$). In order to demonstrate that the time dependence of the RBS-shift is known with sufficient precision, Fig. \ref{fig:residuals}a shows the weighted phase difference after subtraction of all linear terms in the fit model of Eq. (\ref{eqn:fitmodel2}) (including the linear terms from the Taylor expansion of the exponential functions of the RBS-phases). For the 7 measurement runs, an almost parabolic dependence of the residual RBS-shifts (dominated by the quadratic term in the Taylor expansion)  
is the finding. Besides strong variations of the RBS-amplitudes from run to run, a sign change is observed from run ($j=4$) on, i.e., where the magnetic guiding field with respect to its north-south orientation was rotated resulting in slightly different field gradients across the sample cell, i.e., superposition of gradients of the applied field and residual field inside BMSR-2 \cite{Heil,Allmendinger,Schmidt2}. The reduction of these structures by three orders of magnitude (see Fig. \ref{fig:residuals}) indicates that the time structure of the RBS-effects is well under control.
Assuming Gaussian noise, the uncertainty in the phase determination and thus the measurement sensitivity increases according to the Cramer Rao Lower Bound (CRLB) with $\sigma_{\Phi}\propto T^{-\frac{1}{2}}\text{SNR}^{-1}C(T,T_2^*)$ \cite{Gemmel,Heil,Kay}. $T$ is the observation time of coherent spin precession and $C(T,T_2^*)$ is a factor which takes into account the exponential decay of the signal amplitude with the transverse relaxation time $T_{2}^*$ \cite{Gemmel}. The Allan standard deviation (ASD) \cite{Barnes,Gemmel} is the most convenient measure to study the temporal characteristics of our $^{3}$He-$^{129}$Xe co-magnetometer and to identify the power-law model of the phase noise under study. Taking the phase residuals from Fig. \ref{fig:residuals}b, the behavior of the phase uncertainty in the ASD plot is shown in Fig. \ref{fig:ASD}.
Indeed, the observed phase fluctuations decrease as $\propto \tau^{-1/2}$  indicating the presence of white phase noise. After one day of coherent spin precession, the uncertainty in the weighted phase difference is $\sigma_{uncorr}\approx10\mu$rad, typically. The gain in measurement sensitivity compared to the year 2009 measurements mainly arises from two improvements: Firstly, the SNR could be increased by a factor of 4 thanks to the higher xenon polarization of $P_{Xe}\approx 40\%$ and the use of four independent gradiometers. Furthermore, with the larger size of our spherical glass cells, the longitudinal wall relaxation time which scales like $T_{1, wall}\propto R$ could be improved by a factor of 2 for both gas species, i.e., $T_{1, wall}^{He}\approx165$ h and $T_{1, wall}^{Xe}\approx15$ h. In particular for xenon, that resulted in a significant increase of the transverse relaxation time of $T_{2, Xe}^*\approx 8.5$ h compared to $T_{2, Xe}^*\approx 4.5$ h in 2009 \cite{Gemmel}. Thus, coherent spin-precession could be monitored for more than 24 hours ($\approx3\cdot T_{2, Xe}^*$), whereas typical measurement times in 2009 were limited to 14 hours. The longer periods of coherent spin precession had another advantage, inasmuch as the correlated uncertainty which sets the present sensitivity limit of our $^{3}$He-$^{129}$Xe co-magnetometer could be drastically reduced. The big correlated uncertainty ($\sigma_{corr}$) on the sidereal phase modulation is caused by a partly similar time structure of $\Delta\Phi_{\text{c}}(t)$ and the function describing the sidereal phase modulation: Namely, the combined fit to the data of all 7 runs now including the parameterization of the sidereal phase modulation was performed with
\begin{align} \label{eqn:fitmodel3} 
\nonumber \Delta\Phi_{\text{fit}}(t)&=\Delta\Phi_{\text{c}}(t)+ \left\{\begin{array}{ll}\sin\chi_1\cdot e_x\cdot \sin(\Omega_{s}t+\varphi_1)\\
+\sin\chi_1\cdot e_y\cdot\cos(\Omega_{s}t+\varphi_1),  \text{for } j\leq3\\[1ex]
 \sin\chi_2\cdot e_x\cdot\sin(\Omega_{s}t+\varphi_2)\\
+\sin\chi_2\cdot e_y\cdot\cos(\Omega_{s}t+\varphi_2),  \text{for } j\geq 4 \end{array}\right.\\
\end{align}
Here, the suitable choice of coordinate frames and transformations was made as given in Ref. \cite[pp. 7-9]{Kostelecky}. 
With $\chi=\arccos(\cos\Theta \cos\rho)$ we get $\sin\chi_1=0.84$ and $\sin\chi_2=0.98$. The phases were determined to be $\varphi_1=0.103$ and $\varphi_2=-0.677$, using $\varphi=\arctan(-\tan\rho / \sin\Theta)+\varphi_{s}$ \cite{siderealphase}.\\
From the fit, the sidereal phase amplitudes $e_x$ and $e_y$ together with their correlated and uncorrelated uncertainties could finally be extracted to be:
\begin{align}
\label{rawresults}
\nonumber
e_x&= (30\pm 34 \pm 4)\,\mu\text{rad}\\
e_y&= (21\pm 45 \pm 3)\,\mu\text{rad}
\end{align}
The present sensitivity limit of our co-magnetometer is still set by $\sigma_\text{corr}$, which is about a factor of ten higher than $\sigma_\text{uncorr}$. However, compared to the year 2009 measurements, the ratio $\frac{\sigma_{\text{corr}}}{\sigma_{\text{uncorr}}}$ could be reduced by a factor of 5, thanks to the longer periods of coherent spin precession\footnote{The impact of longer spin-coherence times ($T_2^*$) on $\sigma_\text{corr}$ has been studied in \cite[Tab. 1]{Gemmel2}.}. The results of the sidereal phase amplitudes can be expressed in terms of the SME coefficients \cite{Kostelecky,Gemmel2} 
\begin{equation}
\tilde{b}^{\text{n}}_{X,Y}=\frac{1}{2}\frac{\hbar\Omega_s}{\frac{\gamma_{He}}{\gamma_{Xe}}-1}\cdot e_{x,y}~,
\end{equation}
 assuming that the spins and the magnetic moments of the $^{3}$He and $^{129}$Xe nuclei are determined by the valence neutron according to the Schmidt model \cite{Schmidt}\footnote{The use of more accurate nuclear models \cite{Anthony,Dzuba} results in:\\ $\tilde{b}^{\text{n}}_X=(5.1\pm 5.9)\cdot 10^{-34}\text{ GeV}$, $\tilde{b}^{\text{n}}_Y= (3.6\pm 7.8) \cdot 10^{-34}\text{ GeV}$ and $\tilde{b}^{\text{n}}_{\bot}<8.4 \cdot 10^{-34}\text{ GeV}  \text{ (68\% CL)}$.}:
\begin{align}
\label{results1}
\nonumber
\tilde{b}^{\text{n}}_X&=( 4.1\pm 4.7)\cdot 10^{-34}\text{ GeV}\\
\tilde{b}^{\text{n}}_Y&=(2.9\pm 6.2) \cdot 10^{-34}\text{ GeV}~.
\end{align}

\noindent These results can be interpreted as a new upper limit of the equatorial component $\tilde{b}^{\text{n}}_{\bot}$ of the background tensor field interacting with the spin of the bound neutron \cite{explanation}:
\begin{align}
\nonumber
\tilde{b}^{\text{n}}_{\bot}&<6.7 \cdot 10^{-34}\text{ GeV}  &\text{(68\% Confidence Level)}\\
\label{results2b}
\tilde{b}^{\text{n}}_{\bot}&<1.3 \cdot 10^{-33}\text{ GeV}  &\text{(95\% Confidence Level)}
\end{align}
This is an improvement by a factor of 30 compared to our year 2009 measurements and an improvement by a factor of 5 compared to the world's best limit. And there is still room for improvements: Presently, the relatively short $T_{2, Xe}^*$, essentially set by $T_{1, wall}$, limits the total observation time $T$ of free spin-precession. Efforts to increase $T_{1, wall}$ considerably are therefore essential. Besides gain in phase sensitivity according to the $T^{-\frac{1}{2}}$ power law (CRLB), the still dominating correlated uncertainty will approach the uncorrelated one. Furthermore, successive measurement runs can be extended to a period of about 100 days. The long time span gives the possibility to measure an annual variation of a daily sidereal modulation to extract limits on boost-dependent Lorentz and \textit{CPT} violating effects like in Ref. \cite{Cane}.\\[1ex]
We are grateful to our glass blower R. Jera for preparing the low relaxation glass cells from GE180. This work was supported by the Deutsche Forschungsgemeinschaft (DFG) under Contracts No. BA 3605/1-1 and SCHM 2708/3-1, the research center ‘‘Elementary Forces and Mathematical Foundations’’ (EMG) of the University in Mainz, and by PRISMA cluster of excellence at Mainz.


\begin{thebibliography}{99}
\bibitem{siderealphase} At $t_{0, 1}$ the local sidereal time was 2.5 h, which results in a sidereal phase of $\varphi_{s}=0.104\cdot2\pi$.
\bibitem{explanation}
The upper limits for $\tilde{b}^{\text{n}}_{\bot}$ have been calculated according to the following method: 
Let $\mu_t$ be the true but unknown value of $\tilde{b}^{\text{n}}_{\bot}$ and let $r_0$ be the experimentally determined value. The probability density for measuring a value $r$ as a function of $\mu_t$ is called $L(r| \mu_t)$. $L$ can be determined by using the two-dimensional Gaussian distribution of $(\tilde{b}^{\text{n}}_x, \tilde{b}^{\text{n}}_y)$, transforming to polar coordinates $(r,\phi)$ and finally integrating over $\phi$. Let $P(\mu_t| r_0)$ be the probability density of the true value $\mu_t$ as a function of the experimental result $r_0$. Using Bayes' theorem one can calculate $P_1(\mu_t| r_0)$ from $L(x| \mu_t)$. Solving $\int_{0}^{l}P_1(\mu_t| r_0) d\mu_t=0.68$ (resp. $0.95$) gives the appropriate limits $l$.
\end{thebibliography}
\end{document}